\documentclass[preprint]{aastex}

\def\uka { \raisebox{-0.5ex} {\mbox{$\stackrel{<}{\scriptstyle \sim}$}}}

\def\nerr#1#2{{+{#1}}-{#2}}

\slugcomment{Accepted for Publication in the Astrophysical Journal}

\shortauthors{Krawczynski et al.}
\shorttitle{X-Ray and TeV Gamma-Ray Observations of Mrk~421}
\begin{document}
\title{Simultaneous X-Ray and TeV Gamma-Ray Observations of the TeV Blazar Markarian 421
during February and May 2000}
\author{H.~Krawczynski\altaffilmark{1,2,11}, 
R.~Sambruna\altaffilmark{3}, 
A.~Kohnle\altaffilmark{2},
P.S.~Coppi\altaffilmark{1}, 
\\[1ex] The HEGRA Collaboration:
F.~Aharonian\altaffilmark{2}, 
A.~Akhperjanian\altaffilmark{9},
J.~Barrio\altaffilmark{4,5},
K.~Bernl\"ohr\altaffilmark{2},
H.~B\"orst\altaffilmark{7},
H.~Bojahr\altaffilmark{8},
O.~Bolz\altaffilmark{2},
J.~Contreras\altaffilmark{4},
J.~Cortina\altaffilmark{4},
S.~Denninghoff\altaffilmark{4},
V.~Fonseca\altaffilmark{5},
J.~Gonzalez\altaffilmark{5},
N.~G\"otting\altaffilmark{6},
G.~Heinzelmann\altaffilmark{6},
G.~Hermann\altaffilmark{2},
A.~Heusler\altaffilmark{2},
W.~Hofmann\altaffilmark{2},
D.~Horns\altaffilmark{2},
A.~Ibarra\altaffilmark{5},
I.~Jung\altaffilmark{2},
R.~Kankanyan\altaffilmark{2},
M.~Kestel\altaffilmark{4},
J.~Kettler\altaffilmark{2},
A.~Konopelko\altaffilmark{2},
H.~Kornmeyer\altaffilmark{4},
D.~Kranich\altaffilmark{4},
H.~Lampeitl\altaffilmark{2},
E.~Lorenz\altaffilmark{4},
F.~Lucarelli\altaffilmark{5},
N.~Magnussen\altaffilmark{8},
O.~Mang\altaffilmark{7},
H.~Meyer\altaffilmark{8},
R.~Mirzoyan\altaffilmark{4},
A.~Moralejo\altaffilmark{5},
L.~Padilla\altaffilmark{5},
M.~Panter\altaffilmark{2},
R.~Plaga\altaffilmark{5},
A.~Plyasheshnikov\altaffilmark{2,10},
G.~P\"uhlhofer\altaffilmark{2},
G.~Rauterberg\altaffilmark{7},
A.~R\"ohring\altaffilmark{6},
W.~Rhode\altaffilmark{8},
G.~Rowell\altaffilmark{2},
V.~Sahakian\altaffilmark{9},
M.~Samorski\altaffilmark{7},
M.~Schilling\altaffilmark{7},
F.~Schr\"oder\altaffilmark{8},
M.~Siems\altaffilmark{7},
W.~Stamm\altaffilmark{7},
M.~Tluczykont\altaffilmark{6},
H.J.~V\"olk\altaffilmark{2},
C.~A.~Wiedner\altaffilmark{2},
W.~Wittek\altaffilmark{4}}
\altaffiltext{1}{Yale University, P.O. Box 208101, New Haven, CT 06520-8101, USA}
\altaffiltext{2}{Max-Planck-Institut f\"ur Kernphysik,
Postfach 103980, D-69029 Heidelberg, Germany}
\altaffiltext{3}{George Mason University, 4400 University Drive, M/S 3F3, Fairfax, 
VA 22030, USA}
\altaffiltext{4}{Max-Planck-Institut f\"ur Physik, F\"ohringer Ring
6, D-80805 M\"unchen, Germany}
\altaffiltext{5}{Universidad Complutense, Facultad de Ciencias
F\'{\i}sicas, Ciudad Universitaria, E-28040 Madrid, Spain }
\altaffiltext{6}{Universit\"at Hamburg, II. Institut f\"ur
Experimentalphysik, Luruper Chaussee 149,
D-22761 Hamburg, Germany}
\altaffiltext{7}{Universit\"at Kiel, Institut f\"ur Experimentelle und
Angewandte Physik,
Leibnizstra{\ss}e 15-19, D-24118 Kiel, Germany}
\altaffiltext{8}{Universit\"at Wuppertal, Fachbereich Physik,
Gau{\ss}str.20, D-42097 Wuppertal, Germany}
\altaffiltext{9}{Yerevan Physics Institute, Alikhanian Br. 2, 375036
Yerevan, Armenia}
\altaffiltext{10}{On leave from  
Altai State University, Dimitrov Street 66, 656099 Barnaul, Russia}
\altaffiltext{11}{Corresponding author, email krawcz@astro.yale.edu}
\begin{abstract}
In this paper we present the results of simultaneous observations of
the TeV blazar Markarian~421 (Mrk~421) at X-ray and TeV gamma-ray energies
with the {\it Rossi X-Ray Timing Explorer} ({\it RXTE}) and the 
stereoscopic Cherenkov Telescope system of the
HEGRA (High Energy Gamma Ray Astronomy) experiment, respectively.
The source was monitored from February 2nd to February 16th and from
May 3rd to May 8th, 2000. In both energy bands 
several flares with very rapid flux variability were observed.
In the X-ray band, the flux increased and decreased with 
e-folding times as short as about 5~hrs.
The 3-20~keV photon index varied between values of 2.2 and 2.9. 
For 5 pointings the data shows statistically significant 
evidence for spectral curvature.
The photon index varied substantially on very short time scales:
on February 11th it hardened within 1.6 hrs by $\Delta\Gamma=$~0.18 and 
on February 14th it softened within 1.6 hrs by $\Delta\Gamma=$~0.2.
The TeV observations of February 7th/8th showed statistically significant
evidence for substantial TeV flux variability on 30~min time scale.
The TeV energy spectrum averaged over all the observations of the campaign
shows a similar steep slope as in earlier HEGRA observations:
$dN/dE\,=$ $N_0$ $\cdot\, (E/1\,\rm TeV)^{-\Gamma}$ with
$N_0\,=$ $(25\pm 1_{\rm stat})\cdot\,10^{-12}$ 
photons $\rm cm^{-2}$ $\rm s^{-1}$ $\rm TeV^{-1}$
and $\Gamma\,=$ $2.94\,\pm\,0.06_{\rm stat}$. Within
statistical errors no evidence for a curvature of the TeV energy spectrum 
is found.
We show the results of modeling the data with a 
time dependent homogeneous Synchrotron Self-Compton (SSC) model.
The X-ray and TeV gamma-ray emission strengths and energy spectra
together with the rapid flux variability strongly suggest
that the emission volume is approaching the 
observer with a Doppler factor of 50 or higher.
The different flux variability time scales observed at X-rays and TeV 
Gamma-rays indicate that a more detailed analysis will require
inhomogeneous models with several emission zones.
\end{abstract}
\section{Introduction}
\label{intro}
Since its early detection as a source of TeV gamma-rays 
\cite{Punc:92,Petr:96} the BL Lac object Mrk~421 ($z\,=$ 0.031)
has been subject to very intensive studies throughout 
the electromagnetic spectrum.
The study of this extreme gamma-ray loud blazar promises
to elucidate the origin of jets of Active Galactic Nuclei
(AGNs). Furthermore, the source is a laboratory for performing time resolved
studies of the processes of particle acceleration and cooling.
With a luminosity per solid angle of about $10^{44}$~erg s$^{-1}$ sr$^{-1}$  
Mrk~421 is clearly a low-luminosity blazar. 
Nevertheless, the central black hole is estimated to be rather massive:
Gorham et al.\ (1999) estimate a mass of between 1.8$\cdot$ $10^8$ and 
$3.6\,\cdot$ 10$^9$ solar masses.

Mrk~421 has intensively been studied at X-ray energies.
Schubnell (1996) used the pointed X-ray telescopes on board of the {\it RXTE} 
satellite to monitor the source over a time period of 17 days. The 2-10\,keV flux varied 
throughout the campaign by a factor of 10, while the photon index 
showed values between 2.3 to 3.4.
Very detailed observation with integration times of several 
days have been carried through with the BeppoSAX instruments during the 
year 1997 and 1998 (Guainizzi et al.\ 1999; Maliza et al.\ 2000;
Fossati et al.\ 2000a-b).
Photon statistics limited the range in which the spectrum could be determined 
to between 0.1~keV and $\sim15$~keV. While single power law and 
broken power law fits did not describe the data satisfactorily, 
a model of the form $F(E)\,=$ $K\,E^{\alpha_1}\,$ 
$\left[1+(E/E_{\rm b})^f\right]^{(\alpha_1-\alpha_2)/f}$ incorporating continuous 
curvature resulted in statistically acceptable fits to the data \cite{Foss:00b}.
The Spectral Energy Distribution (SED) was found to peak in the 
energy range between $0.1$~keV and 1.1~keV, or below $0.1$~keV
outside the energy range covered by the observations. The spectral slope
around 5~keV was described by photon indices 
between 2.5 and 3.2.

At TeV energies Mrk~421 shows rapid flux variability on time scales down to
a fraction of an hour \cite{Gaid:96}. 
Within statistical errors the Mrk~421 energy spectra measured so far are consistent 
with pure power law spectra. The Whipple collaboration reported 260~GeV - 10~TeV photon indices of
$2.54\pm 0.04_{\rm stat}\pm 0.1_{\rm syst}$ and $2.45\pm 0.1_{\rm stat}\pm
0.1_{\rm syst}$ for two very strong flares measured on May 7th and May
16th, 1996 with integral fluxes above a threshold energy of 350~GeV of 7.4
and 2.8 Crab units, respectively \cite{Kren:99}.
The HEGRA collaboration reported steeper 0.5-5~TeV photon
indices during several medium strong flares during 1997 and 1998
(between 1 and 2 Crab units above 1~TeV)
without statistically significant evidence for a departure 
from the 1997--1998 mean index of 
$3.09\pm0.07_{\rm stat}\pm 0.1_{\rm syst}$ \cite{Ahar:99c}.

The broadband flux variability was studied in several intensive observation campaigns and led
to the discovery of pronounced TeV gamma-ray / X-ray flux correlations \cite{Buck:96,Taka:96}.
Recently, Takahashi et al.\ (2000) combined the results of intensive 
UV and X-ray observations performed with EUVE, BeppoSAX, and ASCA 
in April, 1998 with the TeV lightcurves measured with the CAT, HEGRA, and Whipple
Cherenkov telescopes. The UV/X-ray flux showed ``quasi-periodic'' oscillations 
with a period of approximately 1/2 day and seemed to be well correlated with the TeV flux. 

After observations in early February, 2000
with the HEGRA Cherenkov telescopes and the All Sky Monitor on board
the {\it RXTE} satellite showed an increased TeV gamma-ray and X-ray activity 
of the source at a flux level comparable to the flux of the Crab Nebula, 
we asked the {\it RXTE} GOF to use a fraction of {\it RXTE} AO5 time, 
originally intended for monitoring Mrk~501, to observe the more active source Mrk~421.

In this paper we present the results of the coordinated X-ray ({\it RXTE}) and 
TeV gamma-ray (HEGRA) observations performed from
February 2nd--16th, 2000 and from May 3rd--8th, 2000. 
The rest of the paper is structured as follows.
In Sect.~\ref{obs} we describe the X-ray and TeV gamma-ray data samples and data reduction and
in Sect.~\ref{results} we present the observational results, i.e.\ the X-ray and TeV gamma-ray
lightcurves, the flux correlation properties, the search for the shortest time scales of
flux and spectral variability, and the X-ray and TeV energy spectra.
In Sect.~\ref{disc} we discuss the observational results in the framework of 
Synchrotron Self-Compton (SSC) models.
\section{X-ray and TeV Gamma-Ray Data Sets and Data Reduction}
\label{obs}
\subsection{X-Ray Data}
The X-ray analysis described in the following is based on the 3-20~keV data 
from the Proportional Counter Array (PCA; Jahoda et al. 1996) 
on board the {\it RXTE} satellite. 
We did not use the 15--250 keV data from the 
High-Energy X-ray Timing Experiment (Rothschild et al. 1998)
due to poor signal to noise ratio.
Standard-2 mode PCA data gathered with the top layer of the
operational PCUs (Proportional Counter Units) were analyzed. 
The number of PCUs operational during a pointing
varied between 2 and 5.
After applying the standard screening criteria, 
the net exposure in each Good Time Interval ranged from 16 secs to
3.15 ksecs (see Table~\ref{xflux}). Spectra and lightcurves 
were extracted with \verb+FTOOLSv+5.0. Background models
were generated with the tool \verb+pcabackest+, based on the {\it RXTE} GOF 
calibration files for a ``faint''source (less than 40 counts/sec/PCU).
Response matrices for the PCA data were created with the script \verb+pcarsp+v.2.43.
The spectral analysis was performed with the \verb+XSPEC+v.11.0.1 package.
A constant neutral hydrogen column density of 2~$\rm \cdot~10^{20}~cm^{-2}$ was chosen,
a value which lies close to the 21 cm line HI result
of 1.6~$\rm \cdot~10^{20}~cm^{-2}$ \cite{Dick:90} and the
BeppoSAX spectral absorption result of between $\rm 1.7~\cdot~10^{20}~cm^{-2}$ and
$\rm 3.8~\cdot~10^{20}~cm^{-2}$ \cite{Foss:00b}.
Since the analysis is restricted to the energy region above 3~keV the chosen
hydrogen column density has only a very minor influence on the estimated model parameters.
The majority of measurements were satisfactorily described with single power law models;
for days with long integration times and high count rates single power law models
did not describe the data satisfactorily and we fitted broken power-law
models. The quoted uncertainties on the spectral parameters are on the 67\% confidence
level ($\Delta\chi^2$=1) for the parameters of interest.
\subsection{TeV Gamma-Ray Data}
The TeV gamma-ray analysis presented in this paper is based on observations with the 
HEGRA Cherenkov telescope system (Konopelko et al.\ 1999)
located on the Roque de los Muchachos on the Canary
Island of La Palma (lat.\ 28.8$^\circ$ N, long.\ 17.9$^\circ$ W, 2200
m a.s.l.).  
The observations comprise a total of 61~hrs of best quality data.  
The analysis tools, the procedure of data cleaning
and fine tuning of the Monte Carlo simulations, and the
estimate of the systematic errors on the differential $\gamma$-ray
energy spectra were discussed in detail by Aharonian et al. (1999a,b).
The analysis uses the standard ``loose'' $\gamma$/hadron separation
cuts which minimize systematic errors on flux and spectral estimates
rather than yielding the optimal signal-to-noise ratio.  A software
requirement of two triggered Cherenkov Telescopes within 200~m 
from the shower axis, each with
more than 40 photoelectrons per image and a ``distance'' parameter of
smaller than 1.7$^\circ$ was used.  Additionally, only events with a
minimum stereo angle larger than 20$^\circ$ were admitted to the
analysis.  Integral fluxes for certain energy intervals were
obtained by integrating the differential energy spectra over the 
relevant energy region. 
By this means, the zenith angle dependence of the effective area has been corrected for and
the results are largely independent of the assumed energy spectrum. 
For data runs during which the weather or the
detector performance caused a Cosmic Ray detection rate deviating only
slightly, i.e.\ less than 15\% from the expectation value, the
$\gamma$-ray detection rates and spectra were corrected accordingly.
Spectra and fluxes above an energy threshold of 500~GeV were derived
from the 43~hrs of data from zenith angles smaller than 30$^\circ$. 
The search for variability within individual nights is based on
the integral fluxes above 1~TeV and used all data with zenith angles 
up to 45$^\circ$. 

In the following, only statistical errors will be discussed. The systematic uncertainty in
relative flux values is estimated to be smaller than 5\%. Since the uncertainties in absolute fluxes 
are rather large, i.e.\ 30\%, due to the 15\% uncertainty in absolute energy scale, 
we will quote not only absolute flux values but also the flux strength compared 
to that from the Crab Nebula. 
The HEGRA measurement of the integral flux above 1~TeV from the Crab Nebula 
is $16.7\,\cdot\,10^{-12}$ photons $\rm cm^{-2}$ $\rm s^{-1}$ \cite{Ahar:00b}.
The systematic uncertainty on the 500~GeV~-- 5~TeV photon index is estimated to be 0.1.
\section{Results}
\label{results}
\subsection{Lightcurves}
\begin{figure}[bh]
\begin{center}
\resizebox{12.2cm}{!}{\plotone{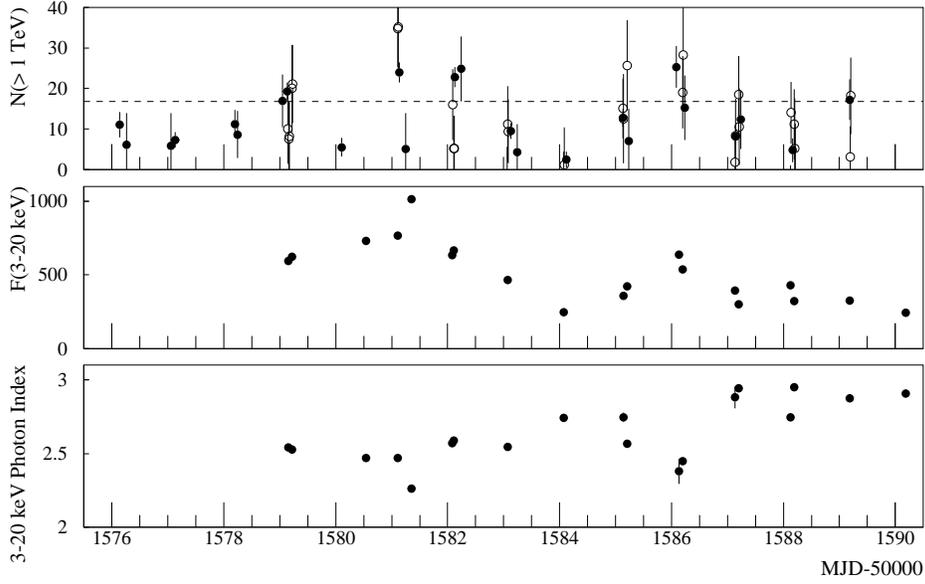}}
\vspace*{-0.4cm}
\end{center}
\caption{\label{lc1} \small 
  The TeV gamma-ray and X-ray results from February, 2000.
  The upper panel shows the integral flux above 1~TeV ($N(>1\,\rm TeV)$)
  in units of 10$^{-12}$ photons $\rm cm^{-2}\,s^{-1}$. The full symbols
  show the diurnal results separately for data from zenith angles below 
  and above 30$^\circ$, the open symbols show the flux as determined 
  with 15 min bins for all bins which overlap with {\it RXTE} observations.
  The dashed line indicates the steady flux level of the Crab Nebula.
  The center-panel shows the 3-20 keV X-ray flux $F(\rm 3-20\,keV)$ 
  in units of $10^{-12}$ $\rm ergs$ $\rm cm^{-2}$ $\rm s^{-1}$.
  The lower panel shows the 3-20~keV photon index.
  MJD 51576 corresponds to February 2nd, 2000.}
\end{figure}

\begin{figure}[bh]
\begin{center}
\resizebox{12.2cm}{!}{\plotone{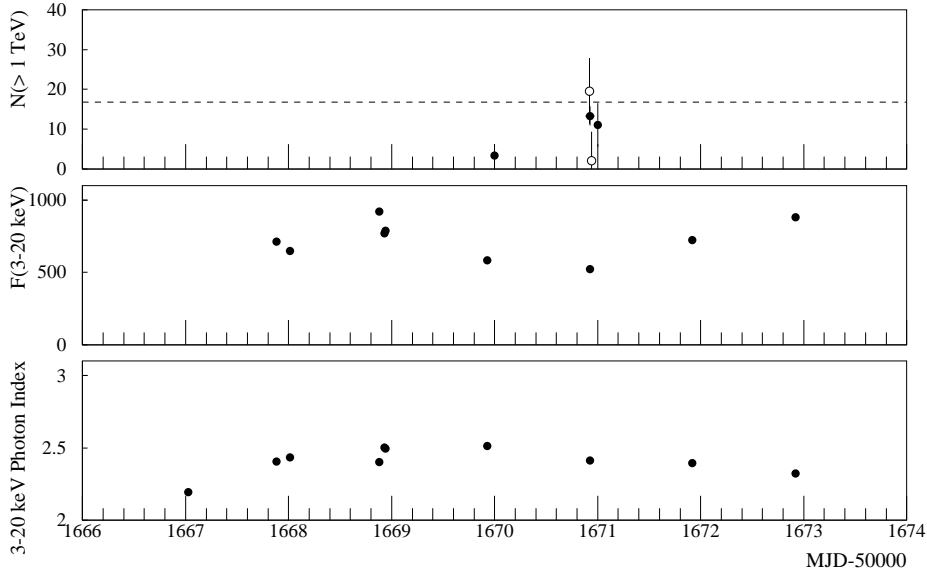}}
\vspace*{-0.4cm}
\end{center}
\caption{\label{lc2} \small 
  The TeV gamma-ray and X-ray results from May, 2000.
  The symbols and units are the same as in Fig.~\ref{lc1}.
  MJD 51667 corresponds to May 3rd, 2000.
}
\vspace*{-7.4cm}
\end{figure}

The TeV gamma-ray and X-ray light curves as well as the 3-20~keV photon indices 
as function of time are shown in Figs.\ \ref{lc1} and \ref{lc2} 
for the February and May observations, respectively. 
The upper panels show the integral photon flux above 1 TeV determined on 
diurnal basis (solid symbols).  The results from observations with zenith angles below 
and above 30$^\circ$ are shown separately to take advantage of the higher 
sensitivity of the Cherenkov Telescope System at zenith angles below 
$\sim$30$^\circ$. 
The diurnal mean integral fluxes vary from values compatible with zero to 
25$\,\cdot\, 10^{-12}$ photons $\rm cm^{-2}$ $\rm s^{-1}$ (1.5 
times the integral flux of the Crab Nebula above 1 TeV). 
Bad weather conditions resulted in only 2 days with good 
TeV data during the May {\it RXTE} observation campaign.

The center-panels of Figs.\ \ref{lc1} and \ref{lc2} show the 3-20~keV X-ray flux. 
During the February and May observations values of between
1.27$\,\cdot\, 10^{-10}$ and 
1.02$\,\cdot\, 10^{-9}$ ergs $\rm cm^{-2}$ $\rm s^{-1}$ were observed.
The TeV gamma-ray and X-ray fluxes are correlated 
in the sense that both fluxes show their maximum emission strength around 
MJD 51581 (February 7th) and MJD 51586 (February 12th) and a minimum around MJD 51584
(February 10th). 
Note that the HEGRA observations had a mean duration of 1.8~h, considerably longer 
than the mean duration of 18~min of the {\it RXTE} observations.
Furthermore, not all HEGRA and {\it RXTE} observations had an overlap in time.
As discussed in the next subsection the TeV Gamma-ray fluxes substantially vary 
on time scales as short as a fraction of an hour.
A more detailed TeV gamma-ray/X-ray flux correlation analysis is 
hampered by the large statistical errors of the TeV flux estimates for 
adequately short integration times.
This is shown by the open symbols in the upper panels of Figs.\ \ref{lc1} and \ref{lc2}. 
Here, the HEGRA flux estimates have been computed with time bins of 15 min duration
and only bins are shown which overlap with {\it RXTE} observations.
Using 15~min time bins, the statistical errors on the TeV flux estimates are 
comparable to the amplitude of the flux variability.

The lower panels in Figs.\ \ref{lc1} and \ref{lc2} show the 3-20~keV photon indices.
The photon indices vary from 2.9 for the days with the lowest flux level
(MJD 51587-51590) to values of 2.2 for the days with the highest flux level 
(MJD 51581, 51667).
The integral TeV gamma-ray fluxes and the results of the power-law fits to the X-ray data 
are summarized in Tables~\ref{tflux} and~\ref{xflux}. 
For 5 observations with good photon statistics the power-law fits do not describe the 
X-ray data satisfactorily; broken power-law fits will be discussed below.
\subsection{Shortest Variability Time Scales}
\begin{figure}[bh]
\begin{center}
\resizebox{12.2cm}{!}{\plotone{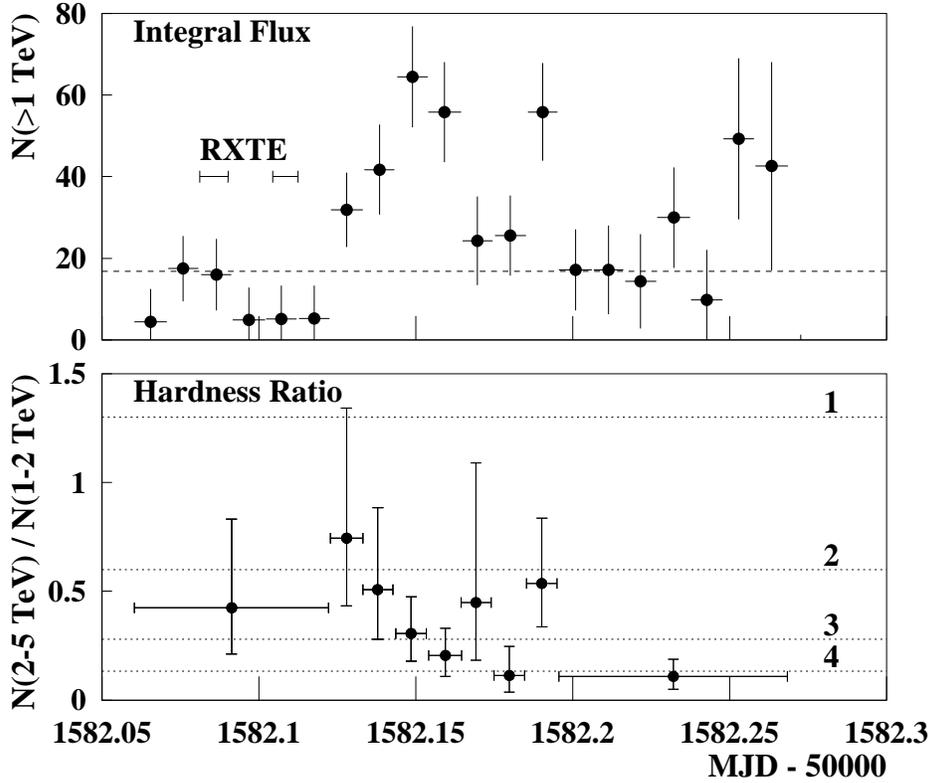}}
\end{center}
\caption{\label{inv} \small 
  The TeV gamma-ray results from February 8th, 2000.
  The upper panel shows the integral flux above 1~TeV ($N(>1\,\rm TeV)$)
  in units of 10$^{-12}$ photons $\rm cm^{-2}\,s^{-1}$ 
  with time bins of 15 min length.
  The dashed line indicates the steady flux level of the Crab Nebula and 
  the 2 horizontal lines show the {\it RXTE} coverage.
  The lower panel shows the $N$(2-5 TeV)/ $N$(1-2 TeV) hardness ratios 
  (median values with 1 sigma confidence intervals). The dotted lines
  show the expected hardness ratios for photon indices of 
  1, 2, 3, and 4, as labelled.
  Statistical errors only, see text for systematic errors.
}
\end{figure}

We performed a search for the shortest TeV flux variability time scale based on
a $\chi^2$-analysis of the integral fluxes above 1~TeV determined with a 15 min binning.
The search revealed one night with significant flux variability. The 
integral fluxes above 1~TeV observed between MJD 51582.06 and MJD 51582.27 (February 8th)
are shown in the upper panel of Fig.~\ref{inv}. Within 1~hr the TeV flux increases from a 
level consistent with zero to $(64\,\pm\,12_{\rm stat})$ $\cdot 10^{-12}$ photons $\rm cm^{-2}$ 
$\rm s^{-1}$ ($3.9\,\pm\,0.7_{\rm stat}$ Crab units). A fit of a constant to the
integral fluxes is rejected with a chance probability of $1.0\cdot10^{-5}$.
Our observation of substantial TeV Gamma-ray flux variability on sub-hour time scale
confirms the existence of substantial sub-hour flux variations 
reported for the strong May 7th, 1996 
Mrk~421 flare \cite{Gaid:96}.

The (2-5 TeV)/(1-2 TeV) hardness ratios are shown in the lower panel of Fig.~\ref{inv}.
The hardness ratios have been computed with the 1-2 TeV and 2-5 TeV photon fluxes
after correction for the (modest) zenith angle dependent variation of the 
effective area over the considered energy range \cite{Ahar:99a}. 
For a TeV energy spectrum of photon index $\Gamma$ the expected hardness 
ratio is then given by 
$r_{\rm exp}(\Gamma)\,=$ $(2^{-\Gamma+1}-5^{-\Gamma+1})\,/\,(1-2^{-\Gamma+1})$ for $\Gamma>1$ and
$\ln{(5/2)}/\ln{(2)}$ for $\Gamma=1$.
The values for $\Gamma\,=$ 1, 2, 3, and~4 are shown as lines in the lower panel
of Fig.~\ref{inv}.
The hardness ratios do not show evidence for spectral variability during the flare. 
A fit of a constant with a mean value of 0.23 gives 
a $\chi^2$ value of 11.7 for 8 degrees of freedom corresponding to a probability of 16.5\%
for a higher value by chance.
The data following MJD 51582.225 has been taken under zenith angles larger than
30$^\circ$, where systematic errors start to be non-negligible. 
For these points (which do not enter strongly the $\chi^2$-value cited above) 
the 1-2 TeV flux is uncertain by about 50\%; accordingly, 
the latest hardness ratio point has a systematic error comparable to the statistical one.
Unfortunately, {\it RXTE} observations were only performed during the first 
1.5 hrs of the 4.8~hrs of HEGRA observations (see Fig.\ \ref{inv}, upper panel)
and did not cover the time of the strong TeV gamma-ray flare.

We analyzed the X-ray flux variability time scale by computing the $e$-folding times
from the flux changes between observations:
$\tau\,=$ $\Delta\,t\,/$ $\Delta\,ln\, F(3-20\rm \, keV)$   
with $\Delta\,t$ the time difference between two observations and 
$\Delta\, ln\, F(3-20\rm \, keV)$  the difference of the 
logarithms of the 3-20~keV fluxes.
The shortest $e$-folding times are given in Table~\ref{inc}.
Flux increases and decreases with $e$-folding times down to $\simeq$5.8~hrs  and 
$\simeq$4.1~hrs have been found, respectively.

We searched for rapid spectral changes by analyzing the change of spectral indices 
between {\it RXTE} observations. 
The fastest spectral changes are listed in Table~\ref{hard}.
The fastest spectral variability is characterized 
by changes in photon index of $\simeq$0.12/hr.
We found similarly rapid spectral hardening as softening: 
on MJD 51585 (February 11th) the spectrum hardened by 0.18 in 1.6 hrs and  
on MJD 51588 (February 14th) the spectrum softened by 0.2 in 1.6 hrs.
\subsection{TeV Gamma-ray and X-ray Energy Spectra}
In the energy range from 500 GeV to 5 TeV the time averaged spectrum of the February 
observations (MJD 51576 -- MJD 51589) is well described by a pure power law model:
$dN/dE\,=$ $N_0$ $\cdot (E/1\,\rm TeV)^{-\Gamma}$ with
$N_0\,=$ $(25\pm 1_{\rm stat})$ photons $\rm cm^{-2}$ $\rm s^{-1}$ $\rm TeV^{-1}$
and $\Gamma\,=$ $2.94\,\pm\,0.06_{\rm stat}$. 
The fit has a $\chi^2$-value of 15.6 for 7 degrees of freedom
(chance probability 5\%).
In Table~\ref{tspec} we give the results of the power law fits for individual days 
for which the accuracy in the determined photon index is better than 0.3 
The photon indices lie between 2.70 and 3.02 but the deviation from the mean index of 2.94 
is statistically not significant. The large $\chi^2$-value of the spectrum of 
MJD 51582.0603 stems from two 
underpopulated bins centered at energies of 700~GeV and 4.35~TeV and 
indicates spectral softening with increasing energy.
A more detailed discussion of the curvature of the TeV energy spectra 
which takes fully into account the systematic uncertainties
is outside the scope of this paper and will be given in an 
upcoming paper in which the full year 2000 data set is included.

As shown in Table~\ref{xflux}, a pure power law fit does not give an
acceptable fit to the data (chance probability well below 1\%)
for 5 {\it RXTE} observations with good photon statistics. 
We find that broken power law models describe the data of these 5
pointings satisfactorily and the results of fits to the 3-25~keV data 
are given in Table \ref{bknpwl}. 
The estimated break energies lie in the range between 6.6~keV and 8~keV and the 
difference between the low and the high energy power law photon indices are about 0.2.
Due to the limited energy coverage of our X-ray observations we did not fit the
data with more complex models incorporating continuous spectral curvature.
We investigated if the other {\it RXTE} data sets are consistent with a similar 
change in spectral index by fitting these data sets with a broken powerlaw model with a
fixed break energy at 7.3~keV 
(the mean break energy found for the data sets of Table \ref{bknpwl}).
Indeed, all fits suggest spectral steepening and we find a median value of 
the change in spectral index of 0.16, very similar to the mean change in 
spectral index of 0.19 found for the data sets of Table \ref{bknpwl}.
\section{Discussion}
\label{disc}
During our observation campaign, the X-ray photon index varied from values between 2.2 to 2.9. 
Cooling of a power law distribution of electrons changes the
synchrotron spectral index by at most 0.5 \cite{Kard:62}. 
Therefore, the observations clearly show that either the spectral index of accelerated 
particles is variable, or that we observe the cooling of electrons near the 
high energy cut-off of the particle acceleration process. 
The TeV gamma-ray/X-ray emission of Mrk~421 is commonly attributed to the SSC mechanism 
(see for alternative models Aharonian et al.\ 2000, and references therein) in which
a population of high energy electrons emits synchrotron radiation at longer 
wavelengths and high energy photons from Inverse Compton (IC) processes of high energy electrons 
with lower energy synchrotron photons at shorter wavelengths. 

Our initial modeling of the X-ray/TeV Gamma-ray data with the time dependent 
SSC code described by Coppi (1992) 
allows already some interesting conclusions which we will detail in the following.
We focus on modeling the observations taken on an individual day, i.e., MJD 51581, where
a large change in X-ray flux and spectrum had been observed and the TeV spectrum has been
determined with reasonable statistical accuracy.
We adopt a spherical emission volume of radius $R=$~2.7$\cdot 10^{15}$~cm
which satisfies the constraints from the observed flux variability
$R\,\uka$ $\delta_{\rm j}$ $c$ $\Delta T_{\rm obs}$ $=\,2.7\cdot 10^{15}$~cm 
for a jet Doppler factor\footnote{
The jet Doppler factor is defined as $\delta_{\rm j}^{-1}\,=\,\Gamma(1-\beta\,\cos{(\theta)})$,
with $\Gamma$ the bulk Lorentz factor, and $\beta$ the bulk velocity in units 
of the speed of light of the emitting volume, respectively, and
$\theta$ is the angle between jet axis and the line of sight as
measured in the observer frame.}
 $\delta_{\rm j}\,=$ 50 and flux variability time scale $\Delta T_{\rm obs}\,=$ 30~min.
We assume a randomly oriented magnetic field of mean 
strength $B\,=$ 0.22~G (in jet frame).
For this magnetic field, electrons of Lorentz factor 
$\gamma_{\rm e}\,=$ $1.8$ $\cdot 10^5$ 
which produce synchrotron radiation with maximum power per 
logarithmic energy band at energy $\varepsilon\,\approx$
$(3/4\pi)$ $\delta_{\rm j}\,h\,e\,<\!\!\sin{\theta}\!\!>\,$$B\,$$\gamma_{\rm e}^2\,$
$(m_{\rm e}\,c)^{-1}\,\approx$ 5~keV
($h$ is Planck's constant, $e$ the electron charge,
$<\!\!\sin{\theta}\!\!>$$=\sqrt{2/3}$) have an observed 
radiative cooling time 
$t_{\rm s}\,=$ $[\frac{4}{3}$ $\sigma_{\mbox{\small T}}\,$$c\,$$\delta_{\rm j}\,$$\frac{B^2}{8\pi\,m_{\rm e}\,c^{2}}\,$$\gamma_{\rm e}$ $]^{-1}~$ 
($\sigma_{\mbox{\tiny T}}$ 
the Thomson cross section) of $\approx 30$ min, comparable to the fastest
variability time scale during the observation campaign.
We inject a power law of accelerated electrons 
$q(\gamma_{\rm e},t)\,=$ {\sl const}
$\cdot\,\gamma_{\rm e}^{-p}\,$ $\exp{(-\gamma_{\rm e}/\gamma_{\rm max}(t))}$
with $p=-2$ (the expected value for diffuse particle acceleration at strong
shocks), and assume a Hubble constant of $H_0\,=\,$$60\,\rm km\,s^{-1}\,Mpc^{-1}$ and
a deceleration parameter of $q_0\,=\,0.5$.

Following Mastichiadis \& Kirk (1997) we model the temporal evolution 
of X-ray flux and spectrum by changing $\gamma_{\rm max}$ only.
We use a damping term for the electron density inside the source of 
$\partial n_{\rm e}/\partial t\,$$\propto\,-n_{\rm e}/t_{\rm esc}$ with
an escape time $t_{\rm esc}$ of 5 light crossing times.
We use a minimum Lorentz factor of accelerated particles of 
$\gamma_{\rm min}\,=\,$$m_{\rm P}/m_{\rm e}\,=\,$ 1836 (ratio of proton to
electron mass), above which diffusive shock acceleration is 
probable to work \cite{Eile:91}.
The results in the X-ray and TeV energy ranges 
do not depend strongly on the value of $\gamma_{\rm min}$ (as long as it is smaller 
than $\approx 10^4$) and the model does not need fine tuning 
of these parameters.
Conversely, the emission strength in the infrared and optical bands do 
depend on the value of $\gamma_{\rm min}$. 
A multiwavelength campaign with observations in these bands, 
together with X-ray and TeV Gamma-ray coverage would make it possible 
to assess this important parameter of particle acceleration theories.

\begin{figure}[bh]
\begin{center}
\resizebox{12.2cm}{!}{\plotone{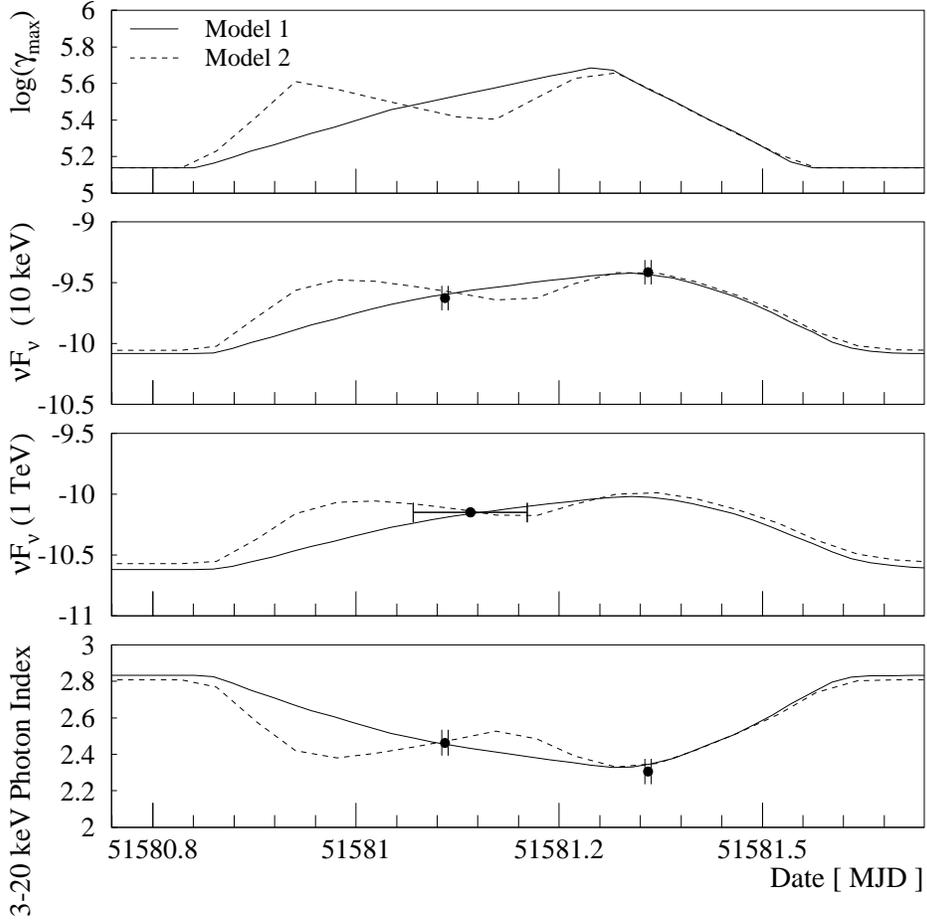}}
\end{center}
\caption{\label{m1} \small 
  From above to below the cut-off electron Lorentz factor, the energy flux at
  10 keV, at 1 TeV, and the 3-20 keV photon indices are shown for the data
  (solid points, horizontal error bars show the length of the observations), 
  and two models (dashed and solid lines). 
  All energy fluxes are in units of $10^{-12}$ erg $\rm cm^{-2}$ $\rm s^{-1}$.
  See text for model parameters and systematic errors on the data points.}
\end{figure}

The solid lines in Fig.~\ref{m1} show the result of calculations
where we modeled the relatively high flux level observed on MJD 51581
by changing $\gamma_{\rm max}$ from an initial value of 
$1.4\,\cdot\,10^5$ to a maximum value of 5.0$\cdot\, 10^5$. 
The model satisfactorily describes the X-ray flux and
X-ray spectral index and the TeV flux.
Note that the observational coverage is much too sparse to
pin down the temporal evolution of the source.
Very different choices of the temporal evolution of the
maximum Lorenz factor of accelerated particles $\gamma_{\rm max}(t)$
are able to describe the data, as shown e.g.\ by the dashed line
in Fig.\ \ref{m1}.
Some properties of the SSC model calculations however do not 
depend strongly on the adopted hypothesis of what causes individual flares.
In the following we will focus on these properties.

\begin{figure}[bh]
\begin{center}
\resizebox{12.2cm}{!}{\plotone{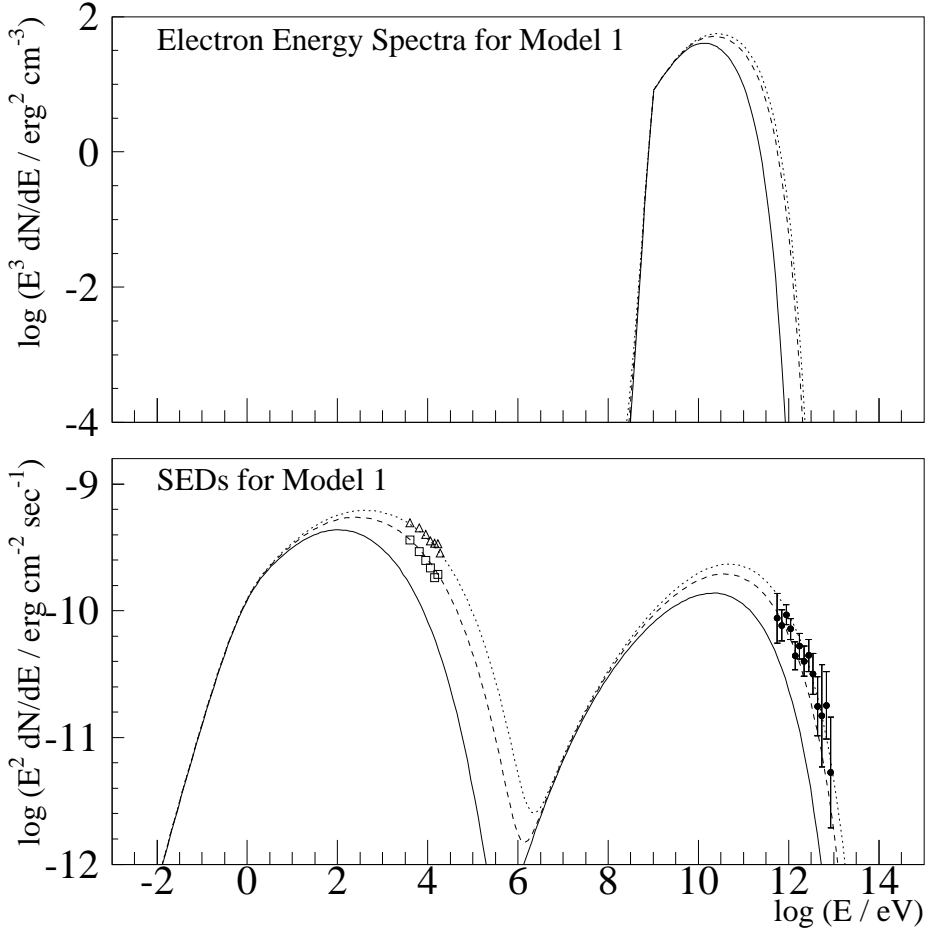}}
\end{center}
\caption{\label{m2} \small 
  The upper panel shows the model estimate for the electron density 
  in the emission region (times the third power of energy), and the lower
  panel shows the observed (symbols) and modelled (lines) SEDs.
  The electron energies are given in the jet frame and the
  photon energies in the observer's frame.
  The model estimates correspond to the solid lines of Fig.~\ref{m1}.
  The squares and the triangles show the {\it RXTE} spectra measured from
  51581.1048-51581.1148 and 51581.3557-51581.3652, respectively.
  The solid circles show the HEGRA data measured during 
  MJD 51581.0702-51581.2119 which included the first {\it RXTE} pointing
  (statistical errors only, see text for systematic errors).
  The solid, dashed and dotted line show the model of the 
  low-flux spectrum before the flare, and during the first and second
  {\it RXTE} pointing, respectively. Note that the model prediction for the
  HEGRA spectrum is to very good approximation shown by the dashed line.
}
\end{figure}

Figure \ref{m2} compares the observed Spectral Energy Distributions (SEDs) 
with the ones from the SSC model 
(``Model 1'' shown by solid lines in Fig.\ \ref{m1}).
Given the constraints on the size of the emission volume from the observed
flux variability, we did not achieve to fit the combined X-ray and TeV Gamma-ray data with
jet Doppler factors substantially below 50. 
Lower Doppler factors resulted either in a strong overproduction of TeV Gamma-rays
for small values of the mean magnetic field ($B\,\ll 0.22$~G), 
or in a too steep TeV energy spectrum for high values of the 
mean magnetic field ($B\gg0.22$~G).
Also a sharper high energy cut-off in the spectrum of accelerated particles did not 
reduce the IC photon yield substantially.
Reducing the jet Doppler factor by using a 
minimum Lorentz factor of accelerated particles requires 
$\gamma_{\rm min}$-values of several times $10^4$ which is difficult
to motivate theoretically.
Taking into account extragalactic extinction would result in even
higher Doppler factors since the effect is expected to
steepen the observed TeV Gamma-ray spectrum while reducing the 
Gamma-ray flux around 1~TeV by a factor $\uka$2 \cite{Steck:98,Prim:01}.
Furthermore, external seed photons, neglected in our analysis, would
result in an even higher model prediction of the emitted TeV Gamma-ray flux.

Our modeling differs from earlier work which indicated Doppler factors 
of about 15 (see e.g.\ Inoue \& Takahara 1996, Mastichiadis \& Kirk 1997, 
Takahashi et al.\ 2000) by one or several of the following reasons:
we use a small emission volume (consistent with flux variability on 30~min time scale),
a minimum Lorentz factor of accelerated electrons well below $10^4$, and
snapshots of an evolving electron spectrum instead of steady state electron
spectra.
Since the cooling times of the electrons responsible for the
X-ray and TeV Gamma-ray emission are comparable to the 
duration of individual flares, steady state electron populations
overestimate the extend to which electron spectra cool before,
during, and after individual flares.
Note that SSC models with Doppler factors of about 15
predict, in agreement with the results of our code,
TeV energy spectra which are softer than the ones 
which have been observed so far
(Inoue \& Takahara (1996), Takahashi et al.\ (2000)).

For high Doppler factors, the energy density of relativistic
electrons and that of the magnetic field are more comparable than for
low Doppler factors (compare Inoue \& Takahara 1996): 
for our choice of $\gamma_{\rm min}$ and $\delta_{\rm j}$ the energy density 
of relativistic particles is $u_{\rm e}\,\approx\,$0.01~erg cm$^{-3}$ 
and that of the magnetic field is $u_{\rm B}\,\approx\,$0.002~erg cm$^{-3}$.
The model requires a modest minimum kinetic power (see e.g.\ Begelman et al.\ 1994)
transported by the jet $L_{\rm j}\,=\,$ $\Gamma^2$ $(u_{\rm e}+u_{\rm P}+u_{\rm B})$ $c$ $r^2$
$\Delta \Omega$ of $\approx$ 4.25$\cdot 10^{43}$ erg s$^{-1}$.
Here we assumed $\Gamma=50$, a distance of the emission region from the central engine of 
$r\,=$ $10^{16}\,$cm, and a solid angle subtended by the jet 
of $\Delta \Omega\,\approx$ $2 \pi(1-cos(\Gamma^{-1}))=$ $1.26\cdot10^{-3}\, \rm sr$.
Furthermore we assumed a factor of $\kappa=1000$ more cold electrons than 
relativistic electrons (the density of relativistic electrons in our models is: 
$\approx\,3$ cm$^{-3}$) 
and an equal number of electrons and cold protons, giving a comoving
energy density in cold protons of $u_{\rm P}\,=$ 
$\rm 4.5$ $\cdot\,((\kappa+1)/1001)$ erg cm$^{-3}$.

In accordance with earlier observations \cite{Gaid:96,Taka:96,Mara:99,Foss:00a}
we find shorter flux variability at TeV energies than at X-ray energies 
with shortest $e$-folding times of $\approx$ 1~hr at TeV energies 
and $\approx$ 5~hrs at X-ray energies, respectively.
This finding could naturally be explained in the framework of 
an inhomogeneous SSC model.
If the region of particle acceleration is relatively small,
an event of enhanced particle acceleration could result in a rapidly variable
TeV Gamma-ray component originating from the vicinity of the acceleration region
while the observed X-rays, dominated by the emission of particles of earlier 
acceleration events, could vary more slowly.
If strong internal shocks accelerate the electrons, one indeed expects that
the accelerated particles are bound to the downstream medium 
by the same scattering processes which enable particle acceleration.
If the density of relativistic particles decreases downstream 
(due to particle diffusion or due to adiabatic expansion of the 
downstream plasma) the SSC mechanism then guarantees that the IC 
emissivity decreases faster than the synchrotron emissivity.
Although the synchrotron emission of such a system has been discussed 
in the literature (see e.g.\ Heavens \& Meisenheimer (1987), 
Kirk et al.\ (1998)), the consequences for the temporal evolution 
and correlation of the synchrotron and the IC components have not 
yet been studied. The upcoming Cherenkov telescope experiments 
CANGAROO~III, H.E.S.S., MAGIC, and VERITAS with one order of 
magnitude higher sensitivity than present instruments will make it 
possible to test such inhomogeneous models, and to infer details about
the geometry and dynamics of the radiating plasma.
To ``map'' in this way the jet at its base in the very vicinity of the 
black hole is a very exciting prospect of Gamma-ray astronomy indeed.
\\[2ex]
{\it Acknowledgments}. 
We thank Jean Swank and the {\it RXTE} GOF for allowing us to use the 
RXTE time for Mrk~421 observations. 
The support of the German ministry for Research and
Technology BMBF and of the Spanish Research Council CYCIT is gratefully
acknowledged. We thank the Instituto de Astrophysica de Canarias
for the use of the site and for supplying excellent working conditions at
La Palma. We gratefully acknowledge the technical support staff of the
Heidelberg, Kiel, Munich, and Yerevan Institutes. 
Rita Sambruna is supported by NASA AOP grant NAG 5-27016.

\onecolumn
\clearpage
%
%
\setlength{\hoffset}{-1.5cm}
\begin{deluxetable}{ccr}
\scriptsize
\tablecaption{Integral TeV fluxes above 1~TeV \protect\newline
(Statistical errors only - see text for systematic errors) \label{tflux}}
\tablewidth{0pt}
\tablehead{
\colhead{Start MJD} & 
\colhead{$t_{\rm obs}\,\tablenotemark{a} \,\,\left[\rm hrs \right]$} & 
\colhead{$N(>\,\rm1\,TeV)$\,\tablenotemark{b}} } 
\startdata
 51576.1020 & 1.92 & 11.02$\pm$3.12 \\
 51576.2404 & 0.98 &  6.08$\pm$7.87 \\
 51577.0476 & 4.02 &  7.34$\pm$1.91 \\
 51577.0433 & 0.79 &  5.89$\pm$8.05 \\
 51578.1634 & 1.61 & 11.24$\pm$3.41 \\
 51578.2354 & 0.67 &  8.64$\pm$5.76 \\
 51579.0422 & 4.16 & 19.29$\pm$2.18 \\
 51579.0067 & 1.89 & 16.90$\pm$6.52 \\
 51580.0390 & 3.21 &  5.50$\pm$2.28 \\
 51581.0702 & 3.40 & 23.96$\pm$2.49 \\
 51581.2272 & 1.16 &  5.12$\pm$8.86 \\
 51582.0603 & 3.63 & 22.79$\pm$2.43 \\
 51582.2240 & 1.16 & 24.81$\pm$7.99 \\
 51583.0502 & 3.77 &  9.50$\pm$2.00 \\
 51583.2213 & 1.17 &  4.34$\pm$6.83 \\
 51584.0460 & 3.43 &  2.43$\pm$1.93 \\
 51585.0683 & 3.27 & 12.70$\pm$2.22 \\
 51585.2158 & 1.17 &  6.98$\pm$7.93 \\
 51586.0680 & 0.93 & 25.32$\pm$5.19 \\
 51586.2136 & 1.16 & 15.21$\pm$7.96 \\
 51587.0840 & 2.72 &  8.22$\pm$2.17 \\
 51587.2103 & 1.19 & 12.31$\pm$7.25 \\
 51588.1250 & 1.85 &  4.75$\pm$2.91 \\
 51589.1660 & 0.87 & 17.20$\pm$5.02 \\
 51669.9837 & 0.63 &  3.41$\pm$2.87 \\
 51670.8837 & 2.14 & 13.33$\pm$2.42 \\
 51670.9810 & 1.02 & 11.03$\pm$5.47 \\
\enddata
\tablenotetext{a}{\hspace*{0.2cm} Net exposure}
\tablenotetext{b}{\hspace*{0.2cm} 
Integral flux above 1~TeV in units of ($10^{-12}$ photons cm$^{-2}$ s$^{-1}$)}
\end{deluxetable}
%
%
\begin{deluxetable}{cccccc}
\scriptsize
\tablecaption{Results of power-law fits to the 3 keV -- 20 keV data
(Statistical errors only) \label{xflux}}
\tablewidth{0pt}
\tablehead{
\colhead{Start MJD} & 
\colhead{$t_{\rm obs}\,\tablenotemark{a} \,\,\left[\rm min \right]$} & 
\colhead{$F_{\rm 3-20\,keV}\,$\tablenotemark{b}} & 
\colhead{$\Gamma\,$\tablenotemark{c}} &
\colhead{$\chi^2_{\rm r} \,/\,\rm d.o.f.\,$\tablenotemark{d}} &
\colhead{$P_{\rm c}\,$\tablenotemark{e}}
}
\startdata
51579.1485 &  18.1 &  385.3 $\pm$2.1 & 2.544 $\pm$ 0.012 & 1.22 / 38 & 0.17\\
51579.2158 &  11.7 &  407.0 $\pm$2.6 & 2.527 $\pm$ 0.014 & 1.46 / 38 &0.033\\
51580.5430 &  52.5 &  496.9 $\pm$1.4 & 2.470 $\pm$ 0.006 & 3.10 / 38 & 4.1$\cdot 10^{-10}$\\
51581.1048 &  14.4 &  521.5 $\pm$3.8 & 2.469 $\pm$ 0.016 & 1.17 / 38 & 0.21\\
51581.3558 &  13.6 &  801.9 $\pm$4.8 & 2.262 $\pm$ 0.012 & 1.31 / 38 &0.097\\
51582.0812 &  13.1 &  402.7 $\pm$2.0 & 2.571 $\pm$ 0.011 & 1.18 / 38 &  0.2\\
51582.1044 &  11.7 &  419.0 $\pm$2.2 & 2.588 $\pm$ 0.011 & 1.48 / 38 &0.029\\
51583.0693 &  24.0 &  300.2 $\pm$1.7 & 2.545 $\pm$ 0.012 & 1.34 / 38 &0.077\\
51584.0736 &  13.3 &  140.9 $\pm$1.7 & 2.743 $\pm$ 0.027 & 0.84 / 38 & 0.75\\
51585.1369 &4.0 &  203.5 $\pm$2.9 & 2.746 $\pm$ 0.031 & 0.71 / 38 & 0.91\\
51585.2035 &  14.7 &  267.9 $\pm$1.7 & 2.568 $\pm$ 0.013 & 0.92 / 38 & 0.61\\
51586.1333 &0.3 &  460.1 $\pm$  18.8 & 2.380 $\pm$ 0.085 & 0.78 / 38 & 0.84\\
51586.2021 &5.3 &  369.4 $\pm$3.8 & 2.450 $\pm$ 0.022 & 0.83 / 38 & 0.76\\
51587.1320 &0.8 &  207.9 $\pm$6.3 & 2.881 $\pm$ 0.072 & 0.77 / 38 & 0.84\\
51587.2017 &  12.3 &  154.2 $\pm$1.7 & 2.942 $\pm$ 0.027 & 0.58 / 38 & 0.98\\
51588.1279 &  12.8 &  245.4 $\pm$2.0 & 2.748 $\pm$ 0.019 & 0.89 / 38 & 0.67\\
51588.1931 &  21.3 &  164.8 $\pm$1.3 & 2.950 $\pm$ 0.019 & 1.08 / 38 & 0.34\\
51589.1891 &  22.1 &  172.2 $\pm$1.4 & 2.875 $\pm$ 0.018 & 1.16 / 38 & 0.23\\
51590.1867 &  21.9 &  126.7 $\pm$1.2 & 2.909 $\pm$ 0.023 & 1.16 / 38 & 0.23\\
51667.0272 &  24.0 & 1018.0 $\pm$1.8 & 2.195 $\pm$ 0.004 & 4.64 / 38 & 9.9 $\cdot10^{-20}$\\
51667.8838 &  22.4 &  506.8 $\pm$2.1 & 2.405 $\pm$ 0.009 & 1.16 / 38 & 0.23\\
51668.0120 &  18.1 &  451.8 $\pm$2.2 & 2.435 $\pm$ 0.010 & 0.79 / 38 & 0.81\\
51668.8814 &  24.0 &  656.7 $\pm$2.2 & 2.401 $\pm$ 0.007 & 1.08 / 38 & 0.34\\
51668.9321 &7.2 &  511.1 $\pm$3.6 & 2.502 $\pm$ 0.015 & 0.89 / 38 & 0.67\\
51668.9406 &  24.8 &  525.2 $\pm$2.0 & 2.495 $\pm$ 0.008 & 1.17 / 38 &    0.22\\
51669.9278 &  49.6 &  385.1 $\pm$1.2 & 2.513 $\pm$ 0.007 & 1.55 / 38 &0.016\\
51670.9250 &  29.1 &  368.8 $\pm$1.1 & 2.412 $\pm$ 0.006 & 1.72 / 38 &  0.004\\
51671.9206 &  28.5 &  517.1 $\pm$1.9 & 2.396 $\pm$ 0.008 & 2.06 / 38 &  1.4$\cdot 10^{-4}$\\
51672.9231 &  11.5 &  664.6 $\pm$2.7 & 2.323 $\pm$ 0.008 & 1.65 / 38 &  0.007\\
\enddata
\tablenotetext{a}{\hspace*{0.2cm} Net exposure}
\tablenotetext{b}{\hspace*{0.2cm} 3-20 keV flux in units of ($10^{-12}$ ergs $\rm cm^{-2}\, s^{-1}$)}
\tablenotetext{c}{\hspace*{0.2cm} 3-20 keV photon index}
\tablenotetext{d}{\hspace*{0.2cm} Reduced $\chi^2$-value and degrees of freedom of the power-law fit}
\tablenotetext{e}{\hspace*{0.2cm} Chance probability for larger reduced $\chi^2$-values}
\end{deluxetable}
%
%
%
%
%
\begin{deluxetable}{cccc}
\scriptsize
\tablecaption{e-folding times of the fastest 3-20 keV flux increases and decreases \label{inc}}
\tablewidth{0pt}
\tablehead{
\colhead{MJD1\tablenotemark{a}} & 
\colhead{MJD2\tablenotemark{b}} & 
\colhead{$\Delta t\,\tablenotemark{c}\,\, \left[ \rm hrs\right]$} & 
\colhead{$\tau \,\tablenotemark{d}\,\,\left[ \rm hrs\right]$} }
\startdata
51585.1369 & 51585.2035   & 1.6    &  5.82  $\pm$  0.32     \\
51586.1333 & 51586.2021   & 1.7    & -7.54  $\pm$  1.45     \\
51587.1320 & 51587.2017   & 1.7    & -5.61  $\pm$  0.61     \\
51588.1279 & 51588.1931   & 1.6    & -4.12  $\pm$  0.12     \\
51668.8814 & 51668.9321   & 1.0    & -4.32  $\pm$  0.14     \\
51668.8814 & 51668.9406   & 1.4    & -6.38  $\pm$  0.14     \\
\enddata
\tablenotetext{a}{\hspace*{0.2cm} Start of first observation}
\tablenotetext{b}{\hspace*{0.2cm} Start of second observation}
\tablenotetext{c}{\hspace*{0.2cm} Time difference between observations}
\tablenotetext{d}{\hspace*{0.2cm} $e$-folding time}
\end{deluxetable}
%
%
\begin{deluxetable}{cccc}
\scriptsize
\tablecaption{Fastest changes of 3-20 keV photon index \label{hard}}
\tablewidth{0pt}
\tablehead{
\colhead{MJD1\tablenotemark{a}} & 
\colhead{MJD2\tablenotemark{b}} & 
\colhead{$\Delta t\,\tablenotemark{c}\,\, \left[ \rm hrs\right]$} & 
\colhead{$\Delta \Gamma \,/\,\Delta t\,\tablenotemark{d}\,\,\left[ \rm hrs^{-1}\right]$} }
\startdata
  51581.1048   &51581.3558   & 6.0   & -0.034 $\pm$ 0.003\\
  51585.1369   &51585.2035   & 1.6   & -0.11  $\pm$ 0.02 \\
  51586.1333   &51587.2017   & 25.6  &  0.022 $\pm$ 0.003\\
  51588.1279   &51588.1931   & 1.6   &  0.12  $\pm$ 0.02 \\
  51668.8814   &51668.9321   & 1.1   &  0.09  $\pm$ 0.02 \\
\enddata
\tablenotetext{a}{\hspace*{0.2cm} Start of first observation}
\tablenotetext{b}{\hspace*{0.2cm} Start of second observation}
\tablenotetext{c}{\hspace*{0.2cm} Time difference between observations}
\tablenotetext{d}{\hspace*{0.2cm} Change in spectral index per 1 hr, negative values denote spectral hardening}
\end{deluxetable}
%
%
\begin{deluxetable}{cccccc}
\scriptsize
\tablecaption{Results of power-law fits to the 500 GeV -- 5 TeV data
(Statistical errors only - see text for systematic errors) \label{tspec}}
\tablewidth{0pt}
\tablehead{
\colhead{Start MJD} & 
\colhead{$t_{\rm obs}\,\tablenotemark{a} \,\,\left[\rm hrs \right]$} & 
\colhead{$N_0\,$\tablenotemark{b}} & 
\colhead{$\Gamma\,$\tablenotemark{c}} &
\colhead{$\chi^2_{\rm r} \,/\,\rm d.o.f.\,$\tablenotemark{d}} &
\colhead{$P_{\rm c}\,$\tablenotemark{e}}
}
\startdata
  51579.0422 &  4.16 & 33.51 \nerr{ 2.76}{ 3.16} & 2.76 \nerr{ 0.16}{ 0.14}&  1.42 / 8 & 0.18\\
  51581.0702 &  3.40 & 44.21 \nerr{ 2.71}{ 3.37} & 2.70 \nerr{ 0.14}{ 0.10}&  0.54 / 8 & 0.83\\
  51582.0603 &  3.63 & 36.27 \nerr{ 2.99}{ 2.76} & 2.72 \nerr{ 0.12}{ 0.10}&  3.05 / 8 
                                                                                       & 0.002\\
  51583.0502 &  3.77 & 19.63 \nerr{ 2.92}{ 3.20} & 2.98 \nerr{ 0.34}{ 0.24}&  0.87 / 8 & 0.54\\
  51585.0683 &  3.27 & 24.90 \nerr{ 3.14}{ 3.22} & 3.02 \nerr{ 0.28}{ 0.24}&  1.09 / 7 & 0.47\\
\enddata
\tablenotetext{a}{\vspace*{-0.5cm}\hspace*{0.2cm} Net exposure}
\tablenotetext{b}{\vspace*{-0.5cm}\hspace*{0.2cm} Normalization constant in units of ($10^{-12}\,\rm photons\,cm^{-2}\, s^{-1}\, TeV^{-1}$)}
\tablenotetext{c}{\vspace*{-0.5cm}\hspace*{0.2cm} Power law photon index}
\tablenotetext{d}{\vspace*{-0.5cm}\hspace*{0.2cm} Reduced $\chi^2$-value and degrees of freedom}
\tablenotetext{e}{\vspace*{-0.5cm}\hspace*{0.2cm} Chance probability for larger reduced $\chi^2$-values}
\end{deluxetable}
\vspace*{-1.5cm}
%
%
\begin{deluxetable}{cccccccc}
\tablecaption{Results of broken power-law fits to the 3 keV -- 25 keV data
(Statistical errors only) \label{bknpwl}}
\tablewidth{0pt}
\tablehead{
\colhead{Start MJD} & 
\colhead{$t_{\rm obs}$ \tablenotemark{a}} & 
\colhead{$k_{\rm 1 \,keV}$\,\tablenotemark{b}} & 
\colhead{$E_{\rm b}\,\,\left[ \rm keV \right]$} & 
\colhead{$\Gamma_1\,$\tablenotemark{c}} &
\colhead{$\Gamma_2\,$\tablenotemark{d}} &
\colhead{$\chi^2_{\rm r} \,/\,\rm d.o.f.\,$\tablenotemark{e}} &
\colhead{$P_{\rm c}\,$\tablenotemark{f}}
}
\footnotesize
\startdata
51580.5430 &  52.5 &   0.383+0.005-0.006&   7.40+0.28-0.32& 2.415+0.009-0.010& 2.646+0.026-0.027&   1.11/46&  0.281\\
51667.0272 &  24.0 &   0.456+0.005-0.005&   6.96+0.31-0.28& 2.137+0.007-0.008& 2.301+0.014-0.012&   1.19/46&  0.173\\
51670.9250 &  29.1 &   0.255+0.005-0.005&   6.60+0.81-0.49& 2.358+0.015-0.014& 2.520+0.034-0.023&   0.67/46&  0.957\\
51671.9206 &  28.5 &   0.353+0.007-0.008&   8.03+1.07-1.03& 2.357+0.013-0.017& 2.560+0.069-0.053&   1.09/46&  0.310\\
51672.9231 &  11.5 &   0.385+0.009-0.009&   7.46+0.84-0.55& 2.268+0.016-0.015& 2.477+0.048-0.035&   0.75/46&  0.897\\         
\enddata
\tablenotetext{a}{\vspace*{-0.5cm}\hspace*{0.2cm} Net exposure in units of (min)}
\tablenotetext{b}{\vspace*{-0.5cm}\hspace*{0.2cm} Flux at 1~keV in units of (photons $\rm keV^{-1}$ $\rm cm^{-2}$ $\rm s^{-1}$)}
\tablenotetext{c}{\vspace*{-0.5cm}\hspace*{0.2cm} Low energy photon index}
\tablenotetext{d}{\vspace*{-0.5cm}\hspace*{0.2cm} High energy photon index}
\tablenotetext{e}{\vspace*{-0.5cm}\hspace*{0.2cm} Reduced $\chi^2$-value and degrees of freedom}
\tablenotetext{f}{\vspace*{-0.5cm}\hspace*{0.2cm} Chance probability for larger reduced $\chi^2$-values}
\end{deluxetable}
\end{document}